\documentclass{ieeetj}
\usepackage{cite}
\usepackage{amsmath,amssymb,amsfonts}
\usepackage{algorithmic}
\usepackage{graphicx,color}
\usepackage{textcomp}
\usepackage{hyperref}
\hypersetup{hidelinks=true}
\usepackage{algorithm,algorithmic}
\def\BibTeX{{\rm B\kern-.05em{\sc i\kern-.025em b}\kern-.08em
    T\kern-.1667em\lower.7ex\hbox{E}\kern-.125emX}}
\AtBeginDocument{\definecolor{tmlcncolor}{cmyk}{0.93,0.59,0.15,0.02}\definecolor{NavyBlue}{RGB}{0,86,125}}

\usepackage{makecell,booktabs,tabularx,adjustbox, pifont,ragged2e,array}
\usepackage[dvipsnames]{xcolor}
\usepackage[most]{tcolorbox}
\usepackage{listings}
\newcommand{\spatialtok}{\texttt{\textcolor[HTML]{E6A7A0}{<|spatial|>}}}
\newcommand{\audiotok}{\texttt{\textcolor[HTML]{BBB7F4}{<|audio|>}}}

\newcommand{\cmark}{\textcolor{Green}{\ding{51}}}
\newcommand{\xmark}{\textcolor{Red}{\ding{55}}}
\newcolumntype{Y}{>{\RaggedRight\arraybackslash}X}
\newcolumntype{C}{>{\Centering\arraybackslash}X}
\renewcommand{\arraystretch}{1.1}

\definecolor{Vivid}{HTML}{9A94F7}
\definecolor{Wheat}{HTML}{F5DEB3} 

\lstdefinestyle{vividAngle}{
	basicstyle=\ttfamily\small,
	breaklines=true,
	columns=fullflexible,
	showstringspaces=false,
	upquote=true,
	moredelim=**[s][\color{Vivid}]{<}{>}
}

\def\OJlogo{}
\def\seclogo{}

\def\authorrefmark#1{\ensuremath{^{\textbf{#1}}}}

\usepackage[table]{xcolor}
\usepackage{array} 
\newcolumntype{g}{>{\columncolor{gray!12}\color{gray!60}}c} 

\definecolor{FIGURECOLOR}{HTML}{1F5A83}
\definecolor{SUBFIGURECOLOR}{HTML}{7C0000}

\newcommand{\subfigcolor}[1]{\textcolor{SUBFIGURECOLOR}{#1}}

\definecolor{baselinegray}{HTML}{9C9C9C}

\begin{document}

\markboth{}{Jiang {et al.}}

\title{Sci-Phi: A Large Language Model Spatial Audio Descriptor}

\author{Xilin Jiang\authorrefmark{1}, Student Member, IEEE, Hannes Gamper\authorrefmark{2}, Member, IEEE\\ and Sebastian Braun\authorrefmark{2}, Senior Member, IEEE}
\affil{Columbia University, New York, NY, USA}
\affil{Microsoft Research, Redmond, WA, USA}
\authornote{Work completed during internship at Microsoft.}

\begin{abstract}
Acoustic scene perception involves describing the type of sounds, their timing, their direction and distance, as well as their loudness and reverberation. While audio language models excel in sound recognition, single-channel input fundamentally limits spatial understanding. This work presents \textit{Sci-Phi}, a spatial audio large language model with dual spatial and spectral encoders that estimates a complete parameter set for all sound sources and the surrounding environment. Learning from over 4,000 hours of synthetic first-order Ambisonics recordings including metadata, \textit{Sci-Phi} enumerates and describes up to four directional sound sources in one pass, alongside non-directional background sounds and room characteristics. We evaluate the model with a permutation-invariant protocol and 15 metrics covering content, location, timing, loudness, and reverberation, and analyze its robustness across source counts, signal-to-noise ratios, reverberation levels, and challenging mixtures of acoustically, spatially, or temporally similar sources. Notably, \textit{Sci-Phi} generalizes to real room impulse responses with only minor performance degradation. Overall, this work establishes the first audio LLM capable of full spatial-scene description, with strong potential for real-world deployment. Demo: https://sci-phi-audio.github.io/demo
\end{abstract}



\maketitle

\section{Introduction}
\label{sec:intro}

A spatial acoustic scene is an organic whole of multiple sound events and ambient noise, together with the environment that shapes them. It includes source identity and content; onsets, offsets, and overlaps; direction and distance; loudness and reverberation; and the room’s overall imprint. These aspects are intertwined, and human listeners naturally bind them into a stable, unified representation: psychophysics studies on human hearing \cite{spatial_hearing, auditory_distance_perception} have shown that perception groups soundscape into coherent auditory objects and scenes using patterns over time and space, with distance and reverberation shaping where sources seem to be and spatial structure enabling selective listening in clutter. To fully analyze and understand an acoustic scene, one needs to detect multiple sources alongside background, track them along time, localize them in azimuth and elevation, estimate distance and level, and characterize the room.

To solve this task, neural network-based machine listeners have progressed along several strands: sound event detection and localization \cite{pann, joint_localize_detection}; automatic speech recognition (ASR) \cite{whisper, mc_asr}; and general-purpose audio understanding with emergent audio large language models (LLMs) \cite{ltu, qwenaudio}. Yet despite strong task performance, these models still fall short of perceiving an acoustic scene as an integrated whole: they typically focus on a single (or dominant) foreground source, omit spatial parameters (e.g., direction and distance), and offer little account of the environment (e.g., reverberation, room volume, noise). This gap motivates us to generalize machine listening from recognizing single auditory objects to narrating entire acoustic scenes. Our research goal is therefore twofold: (i) to investigate whether a machine can understand the entire spatial acoustic scene, including \textit{what}, \textit{when}, \textit{where}, and \textit{how} of the sound sources, and the \textit{environment}, analogous to human perception; and (ii) to build a spatial audio understanding model that can be extended to downstream applications including hearing assistants, robotics perception, navigation, and automatic spatial environment monitoring and annotation.

\begin{table}[t]
	\centering
	\small
	\setlength{\tabcolsep}{4pt}
	\begin{adjustbox}{max width=\linewidth} 
		\begin{tabular}{@{} c| c c c c p{1.4cm} p{1.4cm} p{1.3cm} @{}}
			\toprule
			\textbf{Model} & \textbf{Format} & \textbf{Localization} &
			\textbf{Speech} & \textbf{Audio} &
			\textbf{Noise} & \textbf{Acoustics Params} & \textbf{Test on Real RIR} \\
			\midrule
			BAT \cite{bat} & Binaural & \textcolor{Green}{3D + dist.} & \xmark & \cmark & \xmark & \xmark & \xmark \\
			\emph{Can LLM...?} \cite{can_llm} & FOA & \textcolor{gray}{2D angle} & \textcolor{gray}{en} & \xmark & \xmark & \xmark & \xmark \\
			SING \cite{sing} & Owlet (1ch) & \textcolor{gray}{2D angle} & \textcolor{gray}{en} & \xmark & \xmark & \xmark & \xmark \\
			Phi-4 MM \cite{phi4} & Monaural & \xmark & \textcolor{Green}{8 lang.} & \cmark & \xmark & \xmark & n.a. \\
			\midrule
			\textbf{Sci-Phi} & FOA & \textcolor{Green}{3D + dist.} & \textcolor{Green}{8 lang.} & \cmark & \textcolor{Green}{Type, Loudness} & \textcolor{Green}{Loudness, reverb, room size} & \cmark \\
			\bottomrule
		\end{tabular}
	\end{adjustbox}
	\caption{Existing (spatial) audio LLMs vs.\ \textit{Sci-Phi}, highlighting \textit{Sci-Phi}’s full scene description and generalization ability.}
	\label{tab:existing_models}
\end{table}


This work introduces \textbf{\textit{Sci-Phi}}, \textit{\underline{S}patial-scene \underline{c}omprehension and \underline{i}nference with Phi}, the first spatial audio LLM capable of full spatial-scene description. \textit{Sci-Phi} builds on \texttt{Phi-4 Multimodal} \cite{phi4}, a powerful multimodal LLM for audio understanding and speech recognition that is nevertheless restricted to single-channel audio input. At a high level, \textit{Sci-Phi} couples a \textit{spatial encoder} with an \textit{audio encoder} and is trained to generate scene metadata from \textgreater4,000 hours of synthetic first-order Ambisonics (FOA) mixtures spanning 1–-4 sources, background noise, and diverse rooms. Our contributions are threefold: (i) a spatial audio LLM, integrating a spatial encoder, powered by a spatial data and metadata generation pipeline, for comprehensive spatial-scene description; (ii) a permutation-invariant evaluation protocol with 15 metrics that account for multiple sources and environmental attributes; and (iii) extensive experiments demonstrating generalization to real room impulse responses (RIRs), along with careful analyses across SNR, reverberation, source count and other attributes. \textit{Sci-Phi} advances audio foundation models from isolated object recognition to coherent spatial-scene understanding, with promising results towards real-world generalization.

\section{Related Works}
Research in spatial audio has progressed from sound event detection and localization (SELD) to representation learning and, more recently, to spatial understanding with LLM. Standard SELD systems jointly estimate class labels and locations for multiple sources, with advances in model design \cite{seldnet1, seldnet2}, training objective \cite{multiaccdoa}, and benchmarks \cite{starss23}. Despite strong ad-hoc performance, they assume a small label set and remain tailored to SELD, limiting open-vocabulary and holistic scene understanding. Another line of work explores self-supervised learning (SSL), including contrastive learning \cite{ssl1} and masked reconstruction \cite{ssl2, ssl3}, without the need for labels and therefore learning embeddings that are naturally generalizable to new labels. However, these SSL models were designed and evaluated by task-specific heads for standard SELD and ASR tasks, limiting zero-shot and task co-learning ability. A related direction learns joint spatial audio–text embeddings \cite{devnani2024learning, hu2025salmspatialaudiolanguage} via CLIP/CLAP-like cross-modal contrastive learning \cite{CLIP, CLAP2023}. While useful for captioning and retrieval, these embeddings are generally limited to single sources and are not directly applicable to open-ended text generation, such as multi-source scenery description.

Audio LLMs, monaural or spatial, define today's standard of audio foundation models with open-vocabulary user queries and responses. They typically pair an audio encoder with a pretrained transformer-decoder language model (i.e., GPT \cite{gpt}), leveraging strong linguistic priors and a unified next-token objective across tasks such as SELD, ASR, and more general Q\&A. Table~\ref{tab:existing_models} reviews current spatial audio LLMs to the best of our knowledge; \texttt{Phi-4-Multimodal} \cite{phi4} is included as a representative monaural audio LLM \cite{ltu, qwenaudio}. Although heterogeneous spatial formats and limited open-sourcing hinder direct apples-to-apples comparison, most existing spatial audio LLMs are restricted to one audio domain (speech or non-speech), provide only partial localization (2-D or without distance), and omit background and room acoustics completely. In contrast, \textit{Sci-Phi} offers full spatial-scene understanding with a scalable number of directional sources and is the first to demonstrate generalization on real RIRs.
 
\begin{figure}[tb]
	\centering  
    \includegraphics[width=\linewidth]{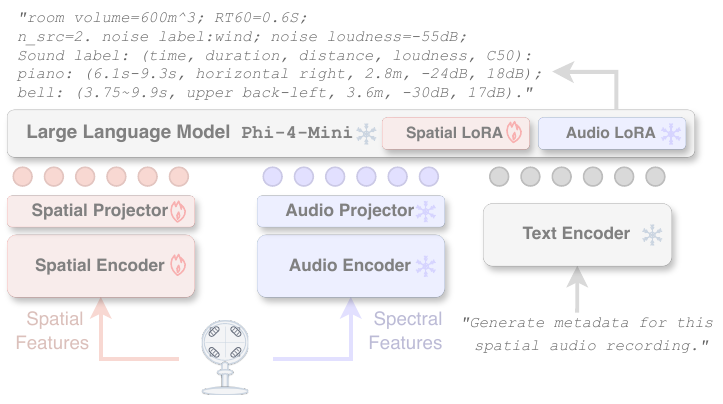}
	\caption{\textit{Sci-Phi} architecture, derived from \texttt{Phi-4-Multimodal} (visual components not shown for clarity). Fire and snowflake mark the trainable and frozen components. Light red, blue, and grey colors correspond to spatial, spectral, and textual features, modules, embeddings, and computation flow.} 
	\label{fig:model}
\end{figure}

\section{Sci-Phi}

\subsection{Multimodal Features and Architecture}
The overall architecture of \textit{Sci-Phi} is shown in Figure~\ref{fig:model}. \textit{Sci-Phi} is a spatial audio LLM with two encoders: a spatial encoder for spatial features and an audio encoder for spectral features. Both features are derived from a first-order Ambisonics (FOA) waveform of four channels $(W,X,Y,Z)$, where $W$ is omnidirectional. Concretely, we compute (i) mel spectrograms of all four channels and (ii) intensity vectors (IVs) \cite{seld_iv} for $(X,Y,Z)$ relative to $W$. These seven maps (4 mel + 3 IV) are stacked as the spatial features, while the spectral features are the mel spectrogram of the $W$ channel alone, since the monaural audio encoder only accepts single-channel inputs.

The spatial encoder borrows the architecture and checkpoint of SELDNet\footnote{Available at https://github.com/partha2409/DCASE2024\_seld\_baseline} \cite{seldnet1, seldnet2}. SELDNet contains 3 convolution layers, 2 gated recurrent units, and 2 self-attention layers. Although the encoder was pretrained on sound event detection and localization, both the amount and coverage of training data, with only 13 sound event labels, and mostly only horizontal spatial direction coverage \cite{starss23}, are insufficient to generalize to more complex acoustic scenes (e.g., our test sets). Therefore, we further finetune the spatial encoder together with the LLM by instruction-tuning on a larger and more diverse training set. While separately pre-training the audio encoder on a larger dataset may help generalization even further, we found the joint training of encoder and LLM to perform well.

We directly use the pretrained monaural audio encoder from \texttt{Phi-4 Multimodal}, which consists of 3 convolution layers and 24 conformer blocks \cite{conformer}. We freeze the monaural audio encoder to maximally preserve its original audio understanding ability trained on monoaural audio, particularly its state-of-the-art speech recognition capability in 8 languages. Finally,  two separate 2-layer linear projectors (trained from scratch for spatial, and frozen for audio encoder) project the spatial and audio encoder outputs to the same dimension (3072) as the text embedding. The spatial, audio, and text embedding are modeled jointly by the LLM \texttt{Phi-4-Mini} \cite{phi4} (3.8B \textit{small} LLM). The LLM reads input in the following format:
\begin{tcolorbox}[
	colback=Wheat!10,        
	colframe=Wheat!50!black, 
	boxrule=0.5pt,
	arc=2mm,                  
	left=6pt,right=6pt,top=0pt,bottom=0pt,
	enhanced,
	breakable                 
	]
	{\ttfamily\scriptsize
		"<|user|>\textbf{\spatialtok}\textbf{\audiotok}<|question|><|end|>\newline<|assistant|><|answer|><|end|>"
	}
\end{tcolorbox}
where \textbf{\spatialtok} and \textbf{\audiotok} are variable-length placeholders for spatial and audio embeddings, respectively.

\subsection{Data Generation}
Because well-annotated spatial audio corpora large enough to train \textit{Sci-Phi} are not available publicly, we synthesize first-order Ambisonics (FOA) training data and paired metadata at scale. Each 10s sample is created by (i) sampling a room with pre-rendered multi-channel room impulse responses (RIRs), (ii) placing 1-–4 directional sound sources distributed in the room and a diffuse background by convolving audio sources with the RIRs, and (iii) randomizing levels, spectral filtering etc.\ before mixing sources and background. The training set contains 1.6 million 10s mixtures ($\sim$4,444h), generated as follows.

\textbf{Rooms and RIRs}. We simulate 10k rooms with the image-source model \cite{allen1979image}. Room sizes range from 4×4×3m$^3$ to 25×25×6m$^3$, and the FOA microphone is placed at a random position. For each room we precompute 64 candidate source positions with a roughly spherically uniform direction distribution. We also record room-level attributes such as reverberation time (RT60) and volume.

\textbf{Sound sources}. As diverse sound source corpora we use speech from CommonVoice \cite{commonvoice} (8 languages, $\sim$385h) and general audio from Freesound ($\sim$230k files) and the BBC sound-effects collection  ($\sim$33k files). We clean tags and captions with an LLM to remove recording-condition notes and sound-irrelevant text, and divide them into single-source and multi-source/ambient files. We use files described as multi-source/ambient as background noises and convolve them with all 64 RIRs from one room to simulate diffuse sound.

\textbf{Metadata and quantization}. Each mixture is accompanied by human-readable scene metadata. Room fields include RT60 and room volume; background fields include noise type and its loudness; each source has a caption (also transcription for speech), onset/offset times, direction, distance, level (dBA), and C50. To stabilize generation and evaluation, we quantize: (i) 3-D direction (azimuth and elevation) into 26 regions using 45$^\circ$ angular bins (e.g., \textit{“upper back-left”, “horizontal front-right”, “above”}), (ii) distance to 0.1m, (iii) RT60 and time to 0.1s, (iv) loudness and C50 to 1dB, and (v) room volume to 100m$^3$. These choices hit a balance between these physical (and mostly continuous) acoustic attributes and simple and descriptive  language targets. The sound levels also allow calculation of SNR.

\textbf{Test sets}. We generate two test sets with 10k clips each (27~h):
a held-out \textbf{\textit{synthetic-RIR}} test set using 100 unseen rooms and unseen audio sources from SoundBible(.com) and speech from VCTK (English only) \cite{yamagishi2019vctk}; a \textbf{\textit{real-RIR}} test set spatializes anechoic sources via real FOA RIRs, and adds real spatial background recordings from 100 real rooms, all from the FOA-MEIR dataset \cite{FOA-MEIR}. The FOA-MEIR datasets contains a set of anechoic sound event recordings, and we again use anechoic English speech from VCTK. The FOA-MEIR test set is limited in spatial coverage: \emph{no sources outside the horizontal plane} (above and below $\pm22.5^\circ$), \emph{no room volume} information, and \emph{only ambient background noise without specific labels}. To test these absent conditions, we have to rely on the synthetic test set only.

\begin{table*}[t]
  \centering
  \caption{We evaluate multiple metrics on multiple sources in arbitrary orders via either (1) per-metric optimal permutations $\mathcal{P}_{\mathrm{OM}}$ or (2) a single $\mathcal{P}_{\mathrm{OS}}$ that maximizes $TupleScore$ (joint \textit{What/When/Where}) with respect to the target sources. Scene-level metrics (global attribute, no permutation) are in gray. $\mathcal{P}_{\mathrm{OS}}$ scores closely match $\mathcal{P}_{\mathrm{OM}}$ per-metric optima. Note: \( \mathcal{P}_{\mathrm{OS}} \) WER may be slightly better than $\mathcal{P}_{\mathrm{OM}}$ because pairs lacking either transcript are skipped.}
  \label{tab:ablation_protocal}
  \setlength{\tabcolsep}{4.5pt}        
  \renewcommand{\arraystretch}{1.15}   
  \resizebox{\textwidth}{!}{%
  \begin{tabular}{c|ggg|g|c|cc|ccc|c|ccc}
    \toprule
    \thead{Protocols} &
    \thead{RoomVol\\ErrLog2} &
    \thead{RT60\\Err (s)} &
    \thead{Noise\\CLAP} &
    \thead{Count\\Accuracy (\%)} &
    \thead{Tuple\\Score} &
    \thead{Source\\CLAP} &
    \thead{WER} &
    \thead{Direction Accuracy\\(\text{XYZ}$\vert$ \text{XY} $\vert$ \text{Z}, \%)} &
    \thead{Zone\\Err ({$^\circ$})} &
    \thead{Distance\\ErrRatio} &
    \thead{Time\\IoU} &
    \thead{Loudness\\Err (dB)} &
    \thead{C50\\Err (dB)} \\
    \midrule
    \multicolumn{14}{l}{\textbf{\textit{On synthetic-RIR test set}}} \\ \hline
    Optimal-Metric $\mathcal{P}_{OM}$ & 0.590 & 0.092 & 0.662 & 91.5 & 0.783 & 0.694 & 0.464 & 85.8 $\vert$ 92.1 $\vert$ 94.0 & 6.1 & 0.228 & 0.815 & 1.011 & 1.217 \\
    \textbf{Optimal-Source $\mathcal{P}_{OS}$} & 0.590 & 0.092 & 0.662 & 91.5 & 0.783 & 0.674 & 0.449 & 82.9 $\vert$ 85.1 $\vert$ 92.0 & 8.4 & 0.258 & 0.802 & 1.253 & 1.348 \\
    \midrule
    \multicolumn{14}{l}{\textbf{\textit{On real-RIR test set}}} \\ \hline
    Optimal-Metric $\mathcal{P}_{OM}$ & u.a. & 0.333 & u.a. & 75.2 & 0.765 & 0.712 & 0.387 & 79.7 $\vert$ 90.1 $\vert$ 88.4 & 10.4 & 0.254 & 0.746 & 1.642 & 1.948 \\
    \textbf{Optimal-Source $\mathcal{P}_{OS}$} & u.a. & 0.333 & u.a. & 75.2 & 0.765 & 0.691 & 0.371 & 77.4 $\vert$ 84.4 $\vert$ 87.5 & 12.0 & 0.292 & 0.737 & 1.975 & 2.203 \\
    \bottomrule
  \end{tabular}%
  }
\end{table*}

\subsection{Training Objective}
\textit{Sci-Phi} is trained to generate a full description of the spatial acoustic scene. We serialize the scene metadata into the \verb+<|answer|>+ string with the template below, starting from the environment to the sources:
\begin{tcolorbox}[
  colback=Wheat!10,
  colframe=Wheat!50!black,
  boxrule=0.5pt,
  arc=2mm,
  left=3pt,right=3pt,top=0pt,bottom=0pt,
  enhanced,
  breakable
]
\begin{lstlisting}[style=vividAngle,basicstyle=\ttfamily\scriptsize,tabsize=4]
room_volume=<room_volume>; RT60=<rt60>;
n_src=<n_src>. noise_label:<noise_type>; 
noise_loudness=<noise_dB>.
Sound label:(time, direction, distance, loudness, C50):
<label_1>: (<time_1>, <direction_1>,
    <distance_1>, <loudness_1>, <C50_1>);
<label_2>: (<time_2>, <direction_2>,
    <distance_2>, <loudness_2>, <C50_2>); 
...
\end{lstlisting}
\end{tcolorbox}
\noindent
Fields in blue are sample-specific parameters. It is important to note that the \textbf{\textit{source enumeration order}} must be fixed a priori for the LLM to learn and will affect its performance (see Table~\ref{tab:ablation_order}). Unless otherwise noted, we order sources by decreasing \textbf{\textit{loudness}}.

The trainable components are the spatial encoder, the spatial projector, and the spatial low-rank adaptation (LoRA) \cite{lora} inside the LLM. We keep the existing mono audio LoRA in \texttt{Phi-4-Multimodal} frozen and initialize the spatial LoRA with the same configuration, i.e., a rank of 320. We optimize the next-token prediction objective below, with $Q$ denoting the question tokens, $A$ the answer tokens, and $X_{\text{spatial}}$, $X_{\text{spectral}}$ the spatial and spectral embeddings:
\begin{align}
	\mathcal{L}= - \sum_{i=1}^{L} \log P_{\theta}\!\left( A_i \,\middle|\, Q,\, X_{\text{spatial}},\, X_{\text{spectral}},\, A_{< i} \right)
\end{align}
\textit{Sci-Phi} and all baseline models were trained for five epochs with an AdamW optimizer \cite{adam}, a total batch size of 24, a peak learning rate of 1.0e-4, a linear learning rate warm-up (5\% steps) followed by linear decay, on 8 NVIDIA A100 GPUs with bfloat16 precision.

\section{Evaluation Method}
While many sentence-level NLP metrics calculate a score between the ground-truth and the generated scene description, they miss precision in specific physical or categorical attributes. Therefore, we extract each attribute (e.g., RT60, direction, distance) from the description and calculate ad-hoc metrics on them. The metrics include the cosine similarity of the audio-aware text embedding from \textbf{CLAP} \cite{CLAP2023} for source and noise descriptions,  \textbf{accuracy} of source counting and direction for which we quantize into 26 XYZ (full sphere), 8 XY (azimuth), and 5 Z (elevation) zones, absolute \textbf{error} of direction (with respect to the center of the quantized zone), RT60 ($s$), loudness ($dB$), C50 ($dB$), and transcription (i.e., \textbf{WER}), or the \textbf{error ratio} or log2 of it for distance ($m$) and room volume ($m^3$), and finally the intersection-over-union (\textbf{IoU}) of estimated vs.\ ground-truth source active intervals. The IoU is defined as
\begin{align}
\mathrm{IoU}(g,s) = \frac{\cap(g,s)}{(t^{g}_{\mathrm{off}}-t^{g}_{\mathrm{on}})
	+ (t^{s}_{\mathrm{off}}-t^{s}_{\mathrm{on}})
	- \cap(g,s)} \label{eq:iou}
\end{align}
where $\cap$ denotes the intersection given by
\begin{equation}
    \cap(g,s) 
	= \max\!\Bigl(0,\ \min(t^{g}_{\mathrm{off}},t^{s}_{\mathrm{off}})
	- \max(t^{g}_{\mathrm{on}},t^{s}_{\mathrm{on}})\Bigr) \label{eq:intersection}
\end{equation}
and $t_\mathrm{on}$ and $t_\mathrm{off}$ denote the onset and offset time of sound events.
A few metrics (room volume, RT60 and background noise type and loudness) are defined and calculated for the entire scene, while others are calculated for each source separately, leading to a critical problem of how to find the best matched sources from the generated description to the solution.

\begin{figure*}[tb]
	\centering  \includegraphics[width=0.98\textwidth]{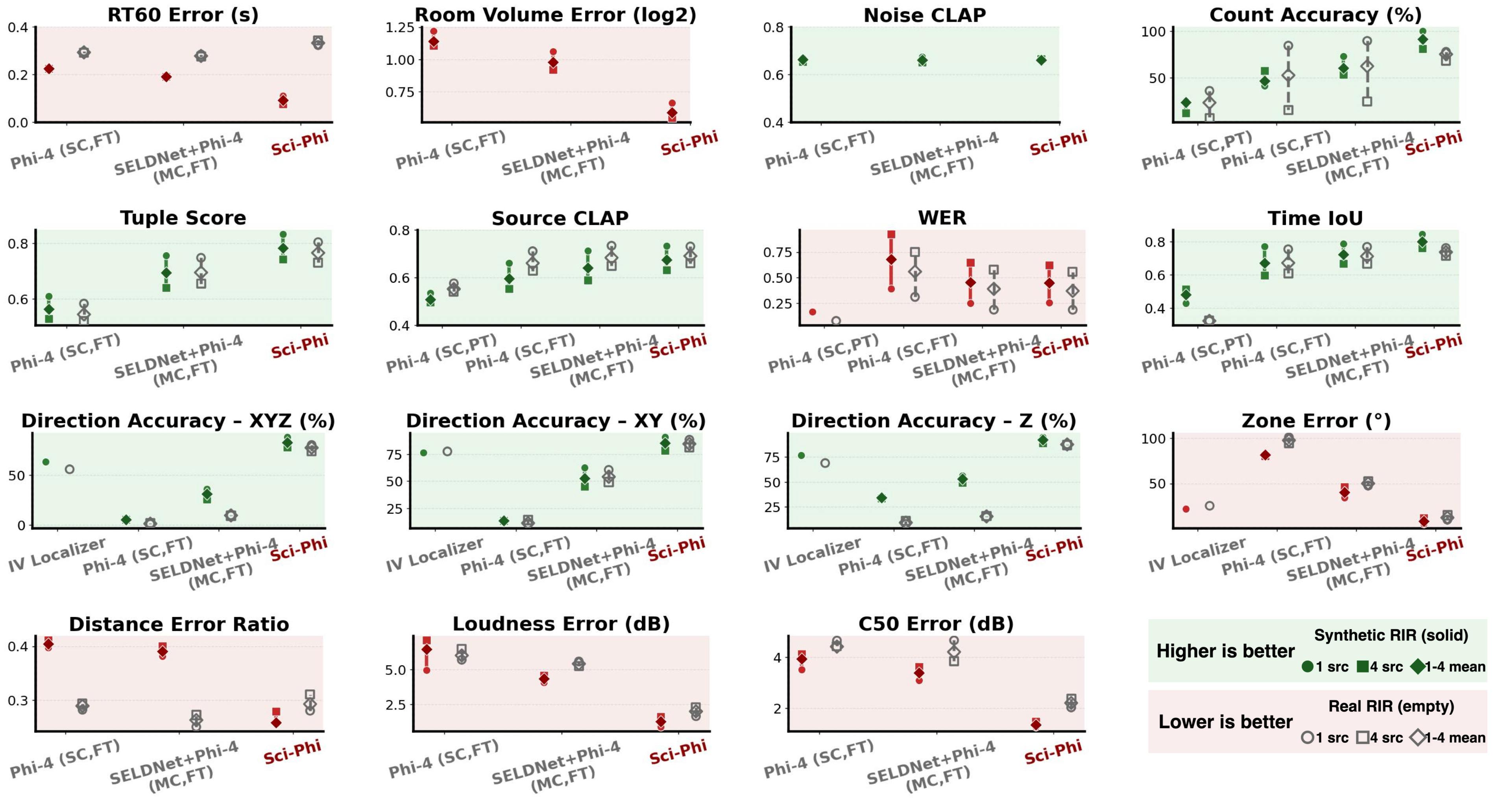}
	\caption{Spatial audio analysis results of the \textbf{synthetic-RIR} (solid) and \textbf{real-RIR} (empty) test sets. Each subplot is one evaluation metric. Green/Red indicates higher/lower is better. Note: room volume error and noise CLAP are missing for real-RIR test due to a lack of ground-truths in FOA-MEIR \cite{FOA-MEIR}.}
	\label{fig:meta_syn_rir}
\end{figure*}

\textbf{Permutation-invariant Evaluation.}
Language models generate tokens autoregressively in a single output stream. While we train \textit{Sci-Phi} to enumerate sources by decreasing loudness, different permutations appear still valid to the human perceivers. Therefore, we argue that evaluation should be \textbf{\textit{order-invariant}} so that correct answers with mismatched or arbitrary orders are not unfairly penalized. We represent each source with six attributes \textit{(label, time, direction, distance, loudness, C50)} and parse both the generated description and the reference into lists of tuples,
$G=[g_1,\dots,g_m]$ and $S=[s_1,\dots,s_n]$. We then seek a permutation matrix $\mathcal{P}$ that reorders $G$ (or $S$) to calculate $\mathrm{Metric}(\mathcal{P}G,S)$ averaged by all sources in the scene.

We could define the \textbf{\textit{optimal-metric permutation}} $\mathcal{P}_{OM} = \max_\text{All $\mathcal{P}$}  \textit{Metric}(\mathcal{P}G,S)$ that maximizes a single metric, like label or direction. However, the downside is that it ignores cross-attribute association: e.g., if
$G=[(\textit{dog},\textit{left}),(\textit{cat},\textit{right})]$ and $S=[(\textit{cat},\textit{left}),(\textit{dog},\textit{right})]$,
per-metric matching can yield perfect scores for both \emph{label} and \emph{direction} despite mismatched association. Instead, we define \textbf{\textit{optimal-source permutation}} $\mathcal{P}_{OS}$ which maximizes a composite \textit{TupleScore} of multiple attributes and does not advantage any particular metric.
\begin{align}
	\mathcal{P}_{OS} = \max_\text{All $\mathcal{P}$}  \textit{TupleScore}(\mathcal{P}G,S)
\end{align}
where the \textit{TupleScore} is geometric mean of \emph{what}, \emph{where}, and \emph{when}, with each term and the final score normalized to $0-1$:
\begin{align}
	\mathit{TupleScore}(g,s)
	&= \bigl(\mathit{What}\!\cdot\!\mathit{Where}\!\cdot\!\mathit{When}(g,s)\bigr)^{1/3} \label{eq:tuple}
\end{align}
with the specific metrics defined as
\begin{align}
	\mathit{What}(g,s) &=
	\begin{cases}
		\dfrac{\mathrm{CLAPScore}(g,s)+1}{2} &\text{for audio}\\
		\max\!\bigl(0,\,1-\mathrm{WER}(g,s)\bigr) &\text{for speech}
	\end{cases} \label{eq:what}\\
	\mathit{Where}(g,s) &= \dfrac{180-\mathrm{ZoneError}(g,s)}{180} \label{eq:where}\\
	\mathit{When}(g,s) &= \mathrm{IoU}(g,s) \label{eq:when}
\end{align}
where the IoU is given in \eqref{eq:iou}. The \textit{What} metric uses \textit{WER} only if a speech source is detected and transcribed in the correct language, and the ground-truth transcription is available.
Otherwise, we default to CLAP score.  

We first determine $\mathcal{P}_{\mathrm{OS}}$ and apply it to compute all metrics. By definition, $\mathcal{P}_{\mathrm{OS}}$ (and its score) equals $\mathcal{P}_{\mathrm{OM}}$ on $TupleScore$;  for other metrics, $\mathcal{P}_{\mathrm{OM}}$ may choose different permutations that define per-metric upper bounds. Nonetheless, in Table~\ref{tab:ablation_protocal}, $\mathcal{P}_{\mathrm{OS}}$ scores closely match $\mathcal{P}_{\mathrm{OM}}$ scores, indicating that \textit{Sci-Phi} learns consistent source-level associations rather than estimating attributes in isolation. At last, our $\mathit{TupleScore}$ and $\mathcal{P}_{\mathrm{OS}}$ formulation can be easily customized for other problems that require finding the best permutation of multiple sources based on multiple metrics.

\section{Results}




As reference points to our proposed system, we include the following baselines:
\begin{enumerate}
  \item \textcolor{baselinegray}{\textbf{IV Localizer}}: ad-hoc DSP intensity-vector localizer.
  \item \textcolor{baselinegray}{\textbf{Phi-4 (SC, PT)}}: pretrained single-channel \texttt{Phi-4-Multimodal} prompted for ASR and multi-sound tagging.
  \item \textcolor{baselinegray}{\textbf{Phi-4 (SC, FT)}}: finetuned single-channel \texttt{Phi-4-Multimodal}.
  \item \textcolor{baselinegray}{\textbf{SELDNet+Phi-4 (MC, FT)}}: finetuned multi-channel \texttt{Phi-4-Multimodal} with a frozen SELDNet encoder.
\end{enumerate}

Figure~\ref{fig:meta_syn_rir} shows the main results for the synthetic RIR test set (solid symbols) and real RIRs (empty symbols). The test data is divided into 4 subsets, depending on the number of directional sources per test clip, i.\,e.\, 1-4 sources. Figure~\ref{fig:meta_syn_rir} marks the results of the subsets of clips with 1 source ($\circ$) and 4 sources ($\square$), and the average over the whole test set with 1--4 sources ($\diamond$).
Across nearly every metric, \textit{Sci-Phi} outperforms all baselines. As expected, the mono Phi-4 baseline fails to localize, and further has poor room volume estimation, source counting, loudness and C50 estimation. Finetuning the mono Phi-4 on our data helps improve counting, source recognition and onset detection.
For synthetic RIRs on average, \textit{Sci-Phi} achieves 91.5\% count accuracy, 0.783 tuple score, 82.9\%, 85.1\%, 92.0\% accuracy for XYZ, XY, Z with $8.4^{\circ}$ zone error, 0.802 time-IoU, and low errors in distance (0.258), loudness (1.25\,dB), and C50 (1.35\,dB). For real RIRs, \textit{Sci-Phi} averages 75.2\% count accuracy, 0.765 tuple score, 77.4\%, 84.4\%, 87.5\% direction accuracy, and 0.737 time-IoU. Reverberation-related metrics degrade more: RT60 error rises from 0.092\,s to 0.333\,s and distance error from 0.258 to 0.292, suggesting potential overfitting to synthetic RIRs. However, other metrics related to semantic, timing, and direction estimation remain strong in unseen real RIRs and sound sources, indicating strong generalization to real rooms.
There are two exceptions aligning with expectations: noise description (CLAP) is on par among all finetuned baselines, implying that predicting diffuse noise type does not demand advanced spatial understanding; for ASR, pretrained single-channel Phi-4 attains lower WER when prompted to only transcribe single speakers, whereas \textit{Sci-Phi}'s unified scene description (including more attributes and potentially up to four speakers) yields slightly higher single-speaker WER, despite beating monaural and spatial baselines finetuned on the same speech and audio corpus. It is interesting that WER is slightly lower for the real RIR test set than for the synthetic test set, although the speech source data is the same in both datasets.

\begin{figure}[!t]
	\centering  \includegraphics[width=\columnwidth]{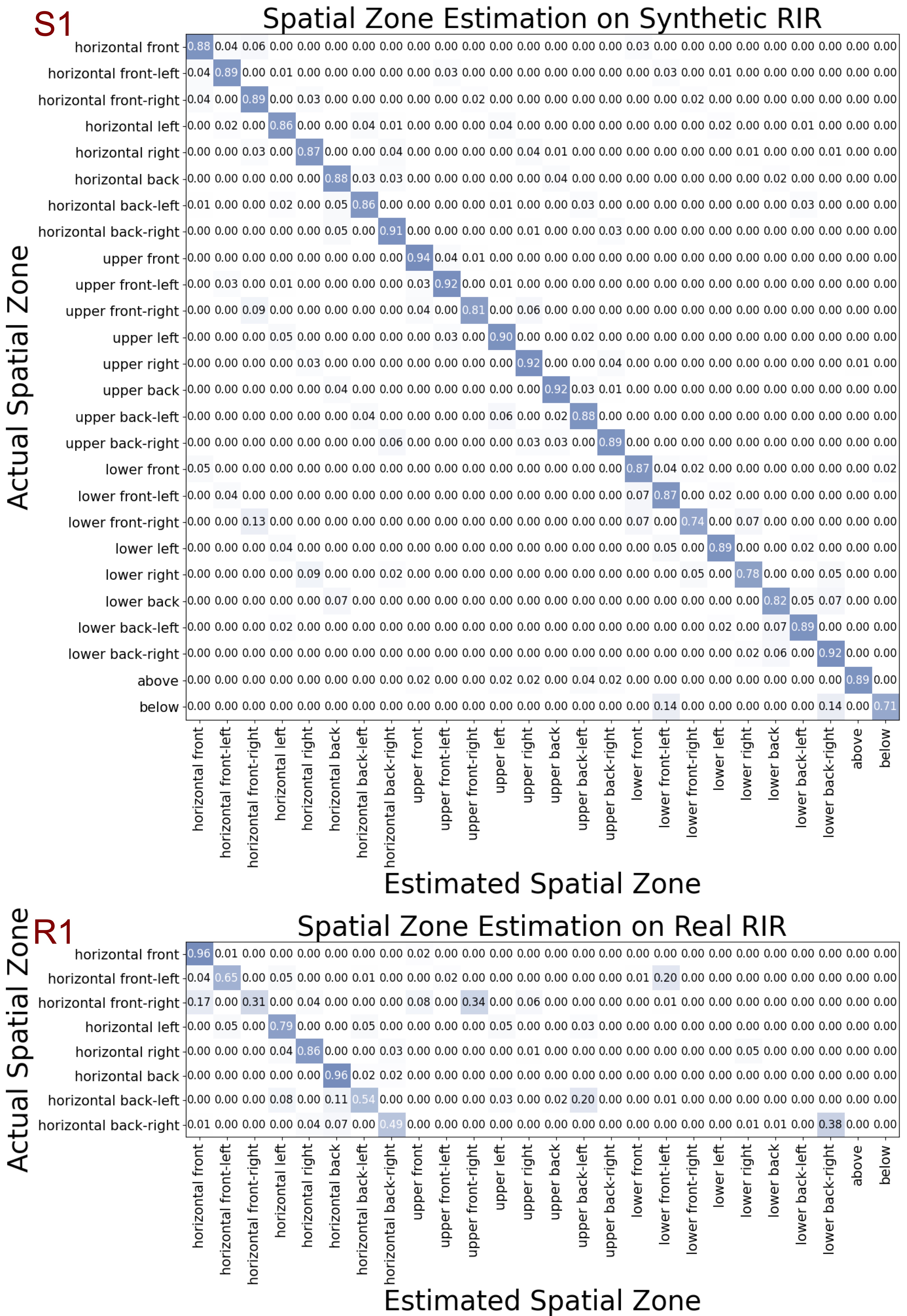}
	\caption{Confusion matrices of single-source localization from the synthetic-RIR (\subfigcolor{\textbf{S1}}) and real-RIR (\subfigcolor{\textbf{R1}}) test sets. Note that we only show a shorter confusion matrix \subfigcolor{\textbf{R1}}, because all source elevations of FOA-MEIR are within $[-22.5^{\circ}, 22.5^{\circ}]$ (\textit{horizontal} label by elevation thresholds), although \textit{Sci-Phi} was trained for and may predict all elevations (\textit{horizontal}, \textit{upper}, or \textit{lower}).}
	\label{fig:confusion_zone}
\end{figure}

\begin{figure}[!t]
	\centering  \includegraphics[width=\columnwidth]{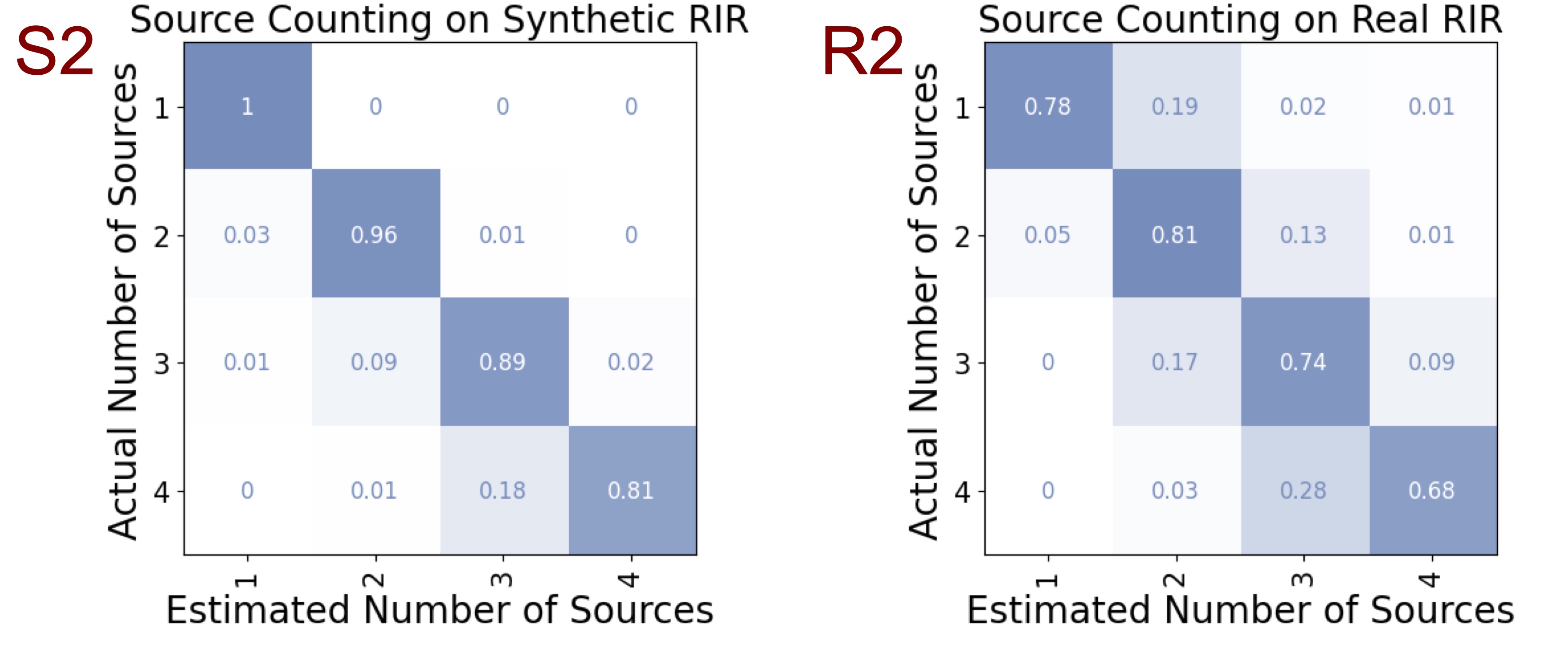}
	\caption{Confusion matrices of source counting from the synthetic-RIR (\subfigcolor{\textbf{S2}}) and real-RIR (\subfigcolor{\textbf{R2}}) test sets.}
	\label{fig:confusion_count}
\end{figure}

\textbf{Localization} Figure~\ref{fig:confusion_zone} \subfigcolor{\textbf{S1}} and \subfigcolor{\textbf{R1}} show the confusion matrix for Sci-Phi on localization for synthetic and real RIRs. We can see that in synthetic RIRs, Sci-Phi localizes sound events with minimal confusion. The confusion matrix for real RIRs is only shown for the horizontal plane, as the test set RIRs from FOA-MEIR contain almost no directions above or below our elevation threshold outside the horizontal zone ($\pm22.5^\circ)$. While there is a mild localization accuracy drop visible compared to synthetic RIRs, we can see that confusions are still rare, and most errors are introduced by confusions into adjacent elevation zones (upper/lower etc.).

\begin{figure}[tb]
  \centering  \includegraphics[width=\columnwidth]{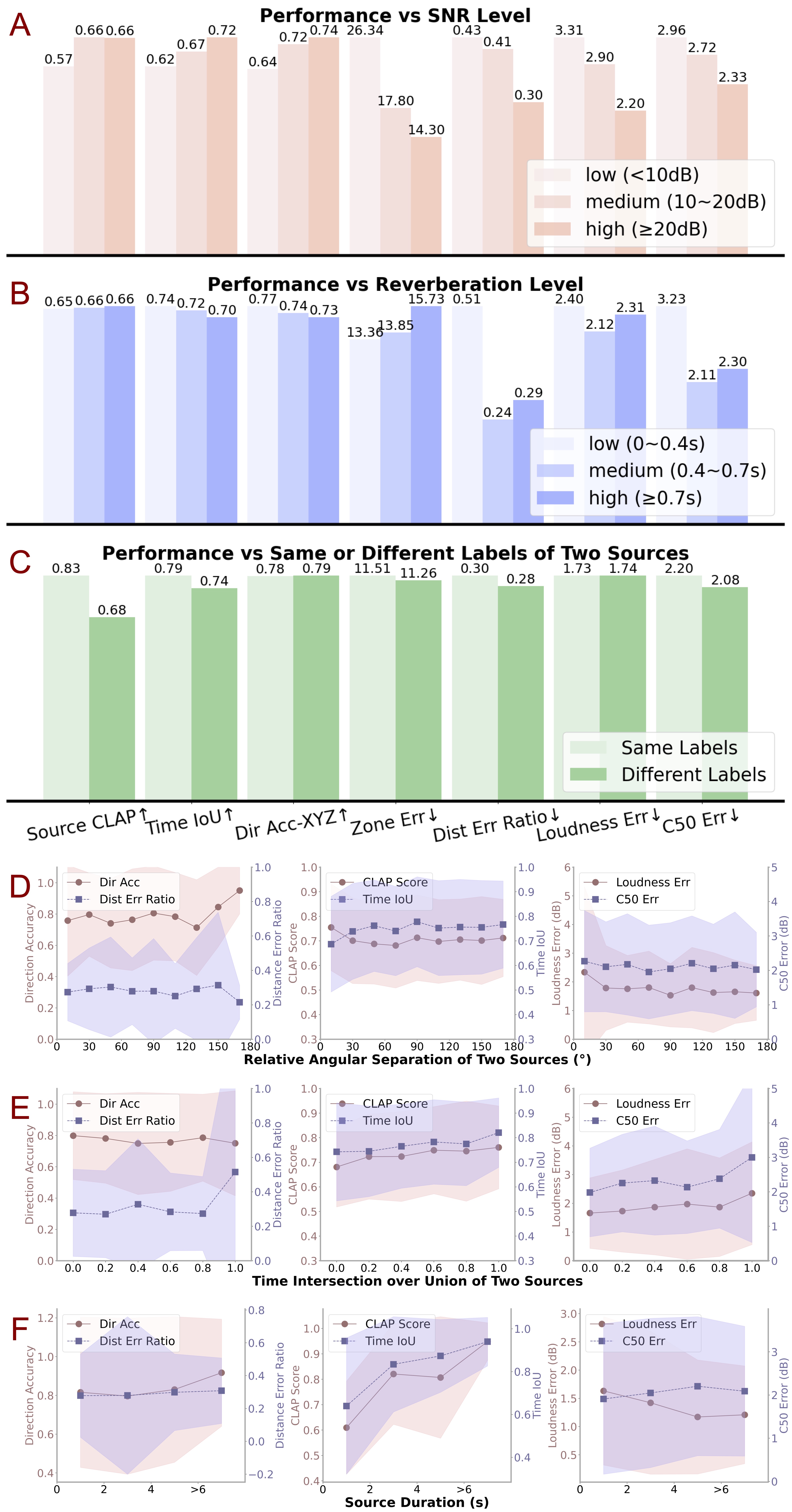}
  \caption{Environmental robustness of \textit{Sci-Phi} on the real-RIR test set. \subfigcolor{\textbf{A}}--\subfigcolor{\textbf{B}}: expected behavior across decreasing SNR and increasing reverberation; \subfigcolor{\textbf{C}}--\subfigcolor{\textbf{E}}: consistent performance even when sources are acoustically, spatially, or temporally close; \subfigcolor{\textbf{F}}: reliable recognition of brief ($\sim1s$) sources, with longer durations providing added context. (The marker position and the shaded area correspond to mean $\pm$ std in \subfigcolor{\textbf{D}}--\subfigcolor{\textbf{F}}.)}
\label{fig:all_analysis}
\end{figure}

\textbf{Source Counting} Figure~\ref{fig:confusion_count} \subfigcolor{\textbf{S2}} and \subfigcolor{\textbf{R2}} shows source counting confusion on synthetic and real RIRs test sets. It is notable that the strongest confusions happen by mis-estimating only $\pm1$ source, while larger errors are rare.
Manual inspection of examples revealed that many source counting errors actually come from two reasons: 1) the model splits a source (coming from a single audio file) into two events, for example a speech file with a pause, or someone kicking a ball into a window, which may be labelled as one event, but is actually two acoustical events (kicking and glass breaking). 2) the model misses a source which can be hardly audible or masked by other sounds because of too low loudness.

\textbf{Environmental Robustness.} We analyze and demonstrate the robustness of \textit{Sci-Phi} under various challenging environments as shown in Figure~\ref{fig:all_analysis}.\subfigcolor{\textbf{A}}--\subfigcolor{\textbf{F}}. As the SNR decreases (\subfigcolor{\textbf{A}}), all metrics, including source recognition (CLAP), temporal alignment (IoU), and localization, degrade \textit{monotonically}. Most metrics likewise worsen as reverberation (RT60) grows (\subfigcolor{\textbf{B}}). A notable exception is \textit{distance}, \textit{loudness}, and \textit{C50} estimation: performance peaks under \textit{moderate} reverberation, consistent with human psychophysics \cite{bradley2003importance} that modest reflections stabilize distance and clarity cues. Even so, direction accuracy and time IoU remain comparatively high, and zone and distance errors increase only mildly for both lower SNR and higher reverberation. 

In \subfigcolor{\textbf{C}}–\subfigcolor{\textbf{E}}, \textit{Sci-Phi} continues to detect two distinct sources even when they are similar in \textit{what} (class label), \textit{where} (direction), or \textit{when} (time interval). Two sources of the exactly same class labels leaves metrics nearly unchanged, and as expected, CLAP scores are higher (better) because predicting the same label to both sources is easier. (\subfigcolor{\textbf{C}}). Reducing angular separation (\subfigcolor{\textbf{D}}) or increasing temporal overlap (\subfigcolor{\textbf{E}}) causes small drops in localization (direction, distance) and acoustic estimates (loudness, C50), yet direction accuracy remains near 80\% and both CLAP and time IoU stay consistent; We only observe a large performance drop when two sources overlap in time almost 100\% (IoU=1). This robustness arises from complementary spatial and spectral (mono) encoders: when two sources are similar along one axis, the other axis can provide discriminative cues. 

Lastly, we examine performance versus source duration in \subfigcolor{\textbf{F}}. \textit{Sci-Phi} is able to recognize short ($\sim1s$) sources, maintaining reliable type (CLAP) and timing (IoU) estimation with competitive direction accuracy, while longer sources naturally provide more context that further lifts these metrics.

\begin{table*}[tb]
	\centering
	\caption{Ablation on the source enumeration order. Results from the synthetic-RIR test set are averaged across 1-4 sources. Only for this ablation, models were trained with 45\% (2k hours) data subset. Environmental metrics are colored in gray, for which the source enumeration order has little effect.} 
	\label{tab:ablation_order}
	\setlength{\tabcolsep}{4.5pt}        
	\renewcommand{\arraystretch}{1.15}   
	\resizebox{\linewidth}{!}{%
		\begin{tabular}{l|ggg|c|c|cc|ccc|c|ccc}
			\toprule
			\thead{Order By} &
             \thead{RoomVol\\ErrLog2} &
             \thead{RT60\\Err (s)} &
             \thead{Noise\\CLAP} &
			\thead{Count\\Accuracy (\%)} &
			\thead{Tuple\\Score} &
			\thead{Source\\CLAP} &
			\thead{WER} &
			\thead{Direction Accuracy\\(\text{XYZ}$\vert$ \text{XY} $\vert$ \text{Z}, \%)} &
			\thead{Zone\\Err ({$^\circ$})} &
			\thead{Distance\\ErrRatio} &
			\thead{Time\\IoU} &
			\thead{Loudness\\Err (dB)} &
			\thead{C50\\Err (dB)} \\
			\midrule
			Zone & 0.639 & 0.104  & 0.667 & \textbf{88.1} & 0.797 & 0.665 & \textbf{0.314} & 77.1 $\vert$ 82.0 $\vert$ 88.5 & 11.1 & 0.279 & 0.802 & 1.587 & 1.606 \\
			Distance & 0.632 & 0.104 & 0.668 & 87.5 & \textbf{0.800} & \textbf{0.666} & 0.323 & 80.7 $\vert$ 83.5 $\vert$ 90.9 & 9.5 & 0.284 & 0.803 & 1.613 & 1.621 \\
			Name & 0.633 & 0.104 & 0.666 & 87.7 & 0.796 & 0.658 & 0.332 & 80.5 $\vert$ 83.9 $\vert$ 90.7 & 9.5 & 0.278 & 0.803 & 1.665 & 1.621 \\
			Onset & 0.632 & 0.105 & 0.670 & 87.7 & \textbf{0.800} & 0.665 & 0.315 & 80.3 $\vert$ \textbf{84.0} $\vert$ 90.6 & 9.7 & 0.278 & \textbf{0.804} & 1.624 & 1.608 \\
			\hline
			\textbf{Loudness} & 0.633 & 0.103 & 0.665 & 87.6 & 0.798 & 0.664 & 0.329 & \textbf{80.8} $\vert$ 83.2 $\vert$ \textbf{91.0} & \textbf{9.3} & \textbf{0.275} & 0.803 & \textbf{1.579} & \textbf{1.579} \\
			\bottomrule
		\end{tabular}%
	}
\end{table*}

\begin{table*}[tb]
	\centering
	\caption{Distinct and combined roles of the spatial and spectral features and encoders. Results from the synthetic-RIR test set are averaged across 1--4 sources.}
	\label{tab:ablation_features}

 \setlength{\tabcolsep}{4.5pt}        
 \renewcommand{\arraystretch}{1.15}   
 \resizebox{\textwidth}{!}{%
 \begin{tabular}{l|ccc|c|c|cc|ccc|c|ccc}
 \toprule
 \thead{Features} &
 \thead{RoomVol\\ErrLog2} &
 \thead{RT60\\Err (s)} &
 \thead{Noise\\CLAP} &
 \thead{Count\\Accuracy (\%)} &
 \thead{Tuple\\Score} &
 \thead{Source\\CLAP} &
 \thead{WER} &
 \thead{Direction Accuracy\\(\text{XYZ}$\vert$ \text{XY} $\vert$ \text{Z}, \%)} &
 \thead{Zone\\Err ({$^\circ$})} &
 \thead{Distance\\ErrRatio} &
 \thead{Time\\IoU} &
 \thead{Loudness\\Err (dB)} &
 \thead{C50\\Err (dB)} \\
 \midrule
 Spatial Only  & \underline{\textbf{0.564}} & \underline{0.097} & \underline{0.647} & \underline{87.1} & \underline{0.699} & \underline{0.553} & 1.189 & \underline{81.7} $\vert$ \underline{84.3} $\vert$ \underline{91.0} & \underline{9.5} & \underline{0.269} & \underline{0.760} & \underline{1.626} & \underline{1.507} \\ 
 Spectral Only & 0.868 & 0.654 & 0.631 & 25.0 & 0.562 & 0.285 & \underline{0.464} & 9.1 $\vert$ 15.8 $\vert$ 46.5 & 75.5 & 0.417 & 0.589 & 7.893 & 6.460 \\ 
 \midrule
 Spatial+Spectral  & 0.590 & \textbf{0.092} & \textbf{0.662} & \textbf{91.5} & \textbf{0.783} & \textbf{0.674} & \textbf{0.449} & \textbf{82.9} $\vert$ \textbf{85.1} $\vert$ \textbf{92.0} & \textbf{8.4} & \textbf{0.258} & \textbf{0.802} & \textbf{1.253} & \textbf{1.348} \\    
 \bottomrule
 \end{tabular}}
\end{table*}

\textbf{Ablations.} Table~\ref{tab:ablation_order} confirms that source enumeration order affects the multi-source description: asking the LLM to order by a given attribute implicitly requires it to be able to estimate that attribute first. Ordering by \textit{loudness} offers the strongest source description performance, yielding the largest number of metric bests. Ordering by zone, distance, or onset is close behind, whereas ordering by name performs worst because open-vocabulary labels introduce ambiguity. Table~\ref{tab:ablation_features} zeros either the spatial or spectral features to ablate on their roles. Spatial features determine counting and localization: with spatial-only, count accuracy stays high (87.1\%) and 3-D direction accuracy remains strong (81.7\%); removing collapses performance toward guessing (25.0\% and 9.1\%). Meanwhile, spectral features determine ASR. Notably, combining both features boosts performance even on metrics where a single feature excels.

\section{Scalability and Extensibility}
\begin{figure*}[tb]
	\centering  \includegraphics[width=\linewidth]{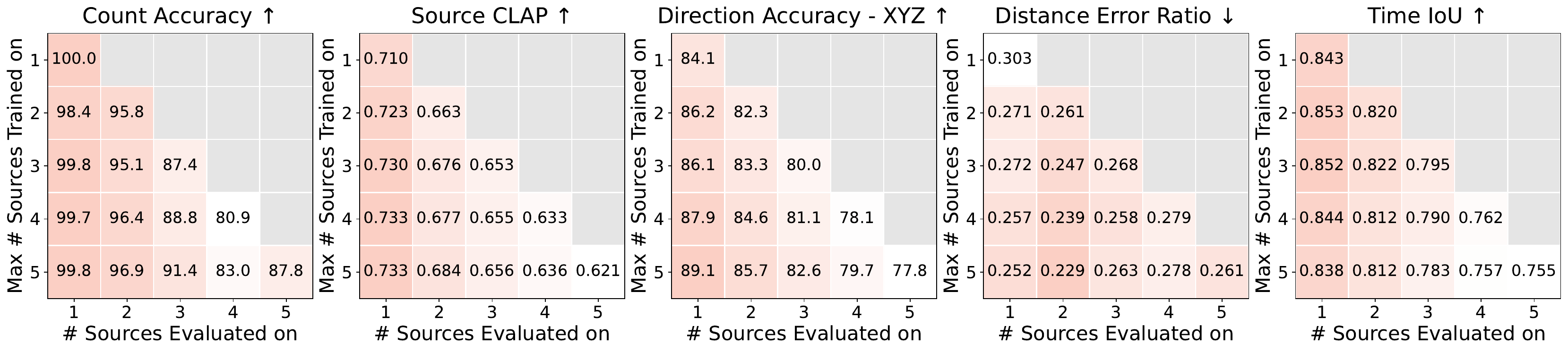}
	\caption{\textit{Sci-Phi} is scalable to the number of sources: still performs strongly up to 5 sources, and preserves performance on fewer sources.}
	\label{fig:more_sources}
\end{figure*}
\textbf{Source Count Scaling.} While we have only reported \textit{Sci-Phi}’s performance up to 4 sources due to only 4 RIR locations per room from the real-RIR (FOA-MEIR) test set, this is \textit{not a ceiling} of the model capability. We additionally generate 400k/1111h 5-source mixtures (in total 2M/5555h of 1--5 source mixtures) and train a new \textit{Sci-Phi} up to 5 sources. In Figure~\ref{fig:more_sources}, we evaluated five model checkpoints trained with a maximum of 1--5 sources on the synthetic-RIR test set. Across all metrics, training on more and evaluating on fewer sources matches or outperforms checkpoints trained only up to that count. With an expected minor performance drop on more sources, \textit{Sci-Phi} still performs strongly up to 5 sources, demonstrating scalability to denser auditory scenes.

\textbf{Spatial Q\&A.} While \textit{Sci-Phi} is trained to output metadata for the entire soundscape; users may want targeted information about a specific source or direction. To support this, we further finetune \textit{Sci-Phi} for one additional epoch with diverse Q\&A prompts, enabling flexible queries such as direction$\rightarrow$label, label$\rightarrow$direction, time$\rightarrow$label, etc. Several examples are provided on our demo page. When multiple or no sources exist in a queried direction, or when a label appears in multiple directions, \textit{Sci-Phi} returns all matches or reports absence.

\section{Conclusion and Limitations}

We introduce \textit{Sci-Phi}, a spatial audio LLM that narrates complete acoustic scenes by jointly modeling spatial and spectral features and generating structured metadata for sources, background, and room acoustics. Trained on synthetic first-order Ambisonics data, \textit{Sci-Phi} generalizes to realistic acoustic conditions with modest performance degradation. Further analysis demonstrates \textit{Sci-Phi}'s robustness and behavior across acoustic conditions such as SNR, reverberation, and the number of directional sound sources. We also propose a consistent evaluation protocol that considers arbitrary ordering of sources and metrics from multiple aspects. This work moves audio foundation models from isolated object recognition toward coherent, multi-attribute spatial scene and object understanding with potential for real-world applications. 

\textbf{Limitations:} Our study has not included systematic evaluation on \textit{in-the-wild} recordings, largely due to the absence of reliable ground truth and mismatches in available label taxonomies. The current framework also assumes stationary sources; generating per-timestep trajectories with an LLM would be computationally very expensive. We leave these directions to future work.


\bibliographystyle{IEEEbib}
\bibliography{refs}

\begin{thebibliography}{10}

\bibitem{spatial_hearing}
Jens Blauert,
\newblock {\em Spatial Hearing: The Psychophysics of Human Sound Localization},
\newblock The MIT Press, 10 1996.

\bibitem{auditory_distance_perception}
Andrew~J. Kolarik, Brian C.~J. Moore, Pavel Zahorik, Silvia Cirstea, and Shahina Pardhan,
\newblock ``Auditory distance perception in humans: a review of cues, development, neuronal bases, and effects of sensory loss,''
\newblock {\em Attention, Perception \& Psychophysics}, vol. 78, pp. 373 -- 395, 2015.

\bibitem{pann}
Qiuqiang Kong, Yin Cao, Turab Iqbal, Yuxuan Wang, Wenwu Wang, and Mark~D Plumbley,
\newblock ``Panns: Large-scale pretrained audio neural networks for audio pattern recognition,''
\newblock {\em IEEE/ACM Transactions on Audio, Speech, and Language Processing}, vol. 28, pp. 2880--2894, 2020.

\bibitem{joint_localize_detection}
Annamaria Mesaros, Sharath Adavanne, Archontis Politis, Toni Heittola, and Tuomas Virtanen,
\newblock ``Joint measurement of localization and detection of sound events,''
\newblock in {\em 2019 IEEE Workshop on Applications of Signal Processing to Audio and Acoustics (WASPAA)}, 2019, pp. 333--337.

\bibitem{whisper}
Alec Radford, Jong~Wook Kim, Tao Xu, Greg Brockman, Christine McLeavey, and Ilya Sutskever,
\newblock ``Robust speech recognition via large-scale weak supervision,''
\newblock in {\em International conference on machine learning}. PMLR, 2023, pp. 28492--28518.

\bibitem{mc_asr}
Tara~N. Sainath, Ron~J. Weiss, Kevin~W. Wilson, Bo~Li, Arun Narayanan, Ehsan Variani, Michiel Bacchiani, Izhak Shafran, Andrew Senior, Kean Chin, Ananya Misra, and Chanwoo Kim,
\newblock ``Multichannel signal processing with deep neural networks for automatic speech recognition,''
\newblock {\em IEEE/ACM Trans. on Audio, Speech, and Language Processing}, vol. 25, no. 5, pp. 965--979, 2017.

\bibitem{ltu}
Yuan Gong, Hongyin Luo, Alexander~H Liu, Leonid Karlinsky, and James Glass,
\newblock ``Listen, think, and understand,''
\newblock in {\em Intl. Conf. on Learning Representations}, 2024.

\bibitem{qwenaudio}
Yunfei Chu, Jin Xu, Xiaohuan Zhou, Qian Yang, Shiliang Zhang, Zhijie Yan, Chang Zhou, and Jingren Zhou,
\newblock ``Qwen-audio: Advancing universal audio understanding via unified large-scale audio-language models,''
\newblock {\em arXiv preprint arXiv:2311.07919}, 2023.

\bibitem{bat}
Zhisheng Zheng, Puyuan Peng, Ziyang Ma, Xie Chen, Eunsol Choi, and David Harwath,
\newblock ``Bat: Learning to reason about spatial sounds with large language models,''
\newblock in {\em International Conference on Machine Learning}, 2024, pp. 61454--61469.

\bibitem{can_llm}
Changli Tang, Wenyi Yu, Guangzhi Sun, Xianzhao Chen, Tian Tan, Wei Li, Jun Zhang, Lu~Lu, Zejun Ma, Yuxuan Wang, et~al.,
\newblock ``Can large language models understand spatial audio?,''
\newblock in {\em Proc. Interspeech 2024}, 2024, pp. 4149--4153.

\bibitem{sing}
Ayushi Mishra, Yang Bai, Priyadarshan Narayanasamy, Nakul Garg, and Nirupam Roy,
\newblock ``Sing: Spatial context in large language model for next-gen wearables,''
\newblock in {\em Forty-second International Conference on Machine Learning}.

\bibitem{phi4}
Abdelrahman Abouelenin, Atabak Ashfaq, Adam Atkinson, Hany Awadalla, Nguyen Bach, Jianmin Bao, Alon Benhaim, Martin Cai, Vishrav Chaudhary, Congcong Chen, et~al.,
\newblock ``Phi-4-mini technical report: Compact yet powerful multimodal language models via mixture-of-loras,''
\newblock {\em arXiv preprint arXiv:2503.01743}, 2025.

\bibitem{seldnet1}
Sharath Adavanne, Archontis Politis, Joonas Nikunen, and Tuomas Virtanen,
\newblock ``Sound event localization and detection of overlapping sources using convolutional recurrent neural networks,''
\newblock {\em IEEE Journal of Selected Topics in Signal Processing}, vol. 13, no. 1, pp. 34--48, 2018.

\bibitem{seldnet2}
Sooyoung Park, Youngho Jeong, and Taejin Lee,
\newblock ``Self-attention mechanism for sound event localization and detection,''
\newblock in {\em DCASE2021 Challenge --- Techn. Reports}, 2021, pp. 1--4,
\newblock Task 3: Sound Event Localization and Detection with Directional Interference.

\bibitem{multiaccdoa}
Kazuki Shimada, Yuichiro Koyama, Shusuke Takahashi, Naoya Takahashi, Emiru Tsunoo, and Yuki Mitsufuji,
\newblock ``Multi-accdoa: Localizing and detecting overlapping sounds from the same class with auxiliary duplicating permutation invariant training,''
\newblock in {\em International Conference on Acoustics, Speech and Signal processing (ICASSP)}. IEEE, 2022, pp. 316--320.

\bibitem{starss23}
Kazuki Shimada, Archontis Politis, Parthasaarathy Sudarsanam, Daniel~A Krause, Kengo Uchida, Sharath Adavanne, Aapo Hakala, Yuichiro Koyama, Naoya Takahashi, Shusuke Takahashi, et~al.,
\newblock ``Starss23: An audio-visual dataset of spatial recordings of real scenes with spatiotemporal annotations of sound events,''
\newblock {\em Adv. in neural information proc. systems}, vol. 36, pp. 72931--72957, 2023.

\bibitem{ssl1}
Xilin Jiang, Cong Han, Yinghao~Aaron Li, and Nima Mesgarani,
\newblock ``Exploring self-supervised contrastive learning of spatial sound event representation,''
\newblock in {\em ICASSP 2024 - 2024 IEEE International Conference on Acoustics, Speech and Signal Processing (ICASSP)}, 2024, pp. 1281--1285.

\bibitem{ssl2}
Antoni Dimitriadis, Siqi Pan, Vidhyasaharan Sethu, and Beena Ahmed,
\newblock ``Spatial hubert: Self-supervised spatial speech representation learning for a single talker from multi-channel audio,''
\newblock {\em arXiv preprint arXiv:2310.10922}, 2023.

\bibitem{ssl3}
Goksenin Yuksel, Marcel van Gerven, and Kiki van~der Heijden,
\newblock ``General-purpose audio representation learning for real-world sound scenes,''
\newblock {\em arXiv preprint arXiv:2506.00934}, 2025.

\bibitem{devnani2024learning}
Bhavika Devnani, Skyler Seto, Zakaria Aldeneh, Alessandro Toso, Elena Menyaylenko, Barry-John Theobald, Jonathan Sheaffer, and Miguel Sarabia,
\newblock ``Learning spatially-aware language and audio embeddings,''
\newblock {\em Adv. in Neural Information Proc. Systems}, vol. 37, pp. 33505--33537, 2024.

\bibitem{hu2025salmspatialaudiolanguage}
Jinbo Hu, Yin Cao, Ming Wu, Feiran Yang, and Jun Yang,
\newblock ``{SALM}: Spatial audio language model with structured embeddings for understanding and editing,'' 2025.

\bibitem{CLIP}
Alec Radford, Jong~Wook Kim, Chris Hallacy, Aditya Ramesh, Gabriel Goh, Sandhini Agarwal, Girish Sastry, Amanda Askell, Pamela Mishkin, Jack Clark, et~al.,
\newblock ``Learning transferable visual models from natural language supervision,''
\newblock in {\em Intl. Conf. on Machine Learning}, 2021, pp. 8748--8763.

\bibitem{CLAP2023}
Benjamin Elizalde, Soham Deshmukh, and Huaming Wang,
\newblock ``Natural language supervision for general-purpose audio representations,''
\newblock in {\em ICASSP 2024-2024 IEEE International Conference on Acoustics, Speech and Signal Processing (ICASSP)}. IEEE, 2024, pp. 336--340.

\bibitem{gpt}
Tom Brown, Benjamin Mann, Nick Ryder, Melanie Subbiah, Jared~D Kaplan, Prafulla Dhariwal, Arvind Neelakantan, Pranav Shyam, Girish Sastry, Amanda Askell, et~al.,
\newblock ``Language models are few-shot learners,''
\newblock {\em Advances in neural information processing systems}, vol. 33, pp. 1877--1901, 2020.

\bibitem{seld_iv}
Masahiro Yasuda, Yuma Koizumi, Shoichiro Saito, Hisashi Uematsu, and Keisuke Imoto,
\newblock ``Sound event localization based on sound intensity vector refined by dnn-based denoising and source separation,''
\newblock in {\em ICASSP 2020-2020 IEEE International Conference on Acoustics, Speech and Signal Processing (ICASSP)}. IEEE, 2020, pp. 651--655.

\bibitem{conformer}
Anmol Gulati, James Qin, Chung-Cheng Chiu, Niki Parmar, Yu~Zhang, Jiahui Yu, Wei Han, Shibo Wang, Zhengdong Zhang, Yonghui Wu, et~al.,
\newblock ``Conformer: Convolution-augmented transformer for speech recognition,''
\newblock in {\em Proc. Interspeech 2020}, 2020, pp. 5036--5040.

\bibitem{allen1979image}
Jont~B Allen and David~A Berkley,
\newblock ``Image method for efficiently simulating small-room acoustics,''
\newblock {\em The Journal of the Acoustical Society of America}, vol. 65, no. 4, pp. 943--950, 1979.

\bibitem{commonvoice}
Rosana Ardila, Megan Branson, Kelly Davis, Michael Kohler, Josh Meyer, Michael Henretty, Reuben Morais, Lindsay Saunders, Francis Tyers, and Gregor Weber,
\newblock ``Common voice: A massively-multilingual speech corpus,''
\newblock in {\em Proceedings of the Twelfth Language Resources and Evaluation Conference}, 2020, pp. 4218--4222.

\bibitem{yamagishi2019vctk}
Junichi Yamagishi, Christophe Veaux, and Kirsten MacDonald,
\newblock ``{CSTR VCTK Corpus}: English multi-speaker corpus for {CSTR} voice cloning toolkit (version 0.92),'' 2019.

\bibitem{FOA-MEIR}
Masahiro Yasuda, Yasunori Ohishi, and Shoichiro Saito,
\newblock ``Echo-aware adaptation of sound event localization and detection in unknown environments,''
\newblock in {\em ICASSP 2022-2022 IEEE International Conference on Acoustics, Speech and Signal Processing (ICASSP)}. IEEE, 2022, pp. 226--230.

\bibitem{lora}
Edward~J Hu, Yelong Shen, Phillip Wallis, Zeyuan Allen-Zhu, Yuanzhi Li, Shean Wang, Lu~Wang, Weizhu Chen, et~al.,
\newblock ``Lora: Low-rank adaptation of large language models.,''
\newblock {\em ICLR}, vol. 1, no. 2, pp. 3, 2022.

\bibitem{adam}
Ilya Loshchilov and Frank Hutter,
\newblock ``Decoupled weight decay regularization,''
\newblock in {\em Intl. Conf. on Learning Representations}.

\bibitem{bradley2003importance}
John~S Bradley, Hiroshi Sato, and Michel Picard,
\newblock ``On the importance of early reflections for speech in rooms,''
\newblock {\em The Journal of the Acoust. Soc. of America}, vol. 113, no. 6, pp. 3233--3244, 2003.

\end{thebibliography}

\vfill\pagebreak

\end{document}